\documentclass[thmsa]{article}
\usepackage{amssymb}
\usepackage{amsmath,amsthm,amsfonts}  
\pdfoutput=1
\usepackage[applemac]{inputenc}  \usepackage{graphicx}
\usepackage{amssymb}
\usepackage{amsmath}
\usepackage{amsthm}
\usepackage{pdfsync}
\usepackage{fancyhdr}
\usepackage[applemac]{inputenc}

\usepackage{color}
\usepackage{xcolor}
\usepackage[breaklinks,colorlinks,backref]{hyperref}
\hypersetup{
    colorlinks, 
    linktoc=all, 
    linkcolor=red,  
  citecolor=hyptxt,
  urlcolor=blue}
  \hypersetup{
  citebordercolor=,
  filebordercolor=green,
  linkbordercolor=blue
}
\definecolor{hyptxt}{rgb}{0.7, 0.4, 0.9}
\usepackage{bibtopic}

\newenvironment{remark}[1][Remark]{\begin{trivlist}
\item[\hskip \labelsep {\bfseries #1}]}{\end{trivlist}}
\newenvironment{conclusion}[1][Conclusion]{\begin{trivlist}
\item[\hskip \labelsep {\bfseries #1}]}{\end{trivlist}}

\renewcommand{\qed}{\hfill $\square$}

\newcommand{\ket}[1]{|\kern.3ex#1\kern.3ex\rangle}
\newcommand{\bra}[1]{\langle\kern.3ex #1 \kern.3ex|}
\newcommand{\scalar}[2]{\langle\kern.3ex #1 \kern.3ex|\kern.3ex#2\kern.3ex\rangle}

\newcommand{\NN}{\mathcal{N}}

\def\N{\mathbb{N}}

\def\deq{\stackrel{\mathrm{def}}{=}}

\def\ud{\mathrm{d}}
\definecolor{hervecolor}{rgb}{0.8,0,0.7}

\numberwithin{equation}{section}

\begin{document}

\title{Entropies of deformed binomial distributions}
\author{
H. Bergeron$^{\mathrm{a}}$, 
E.M.F.  Curado$^{\mathrm{b,c}}$,  \\
J.P. Gazeau$^{\mathrm{b,d}}$, 
 Ligia M.C.S. Rodrigues$^{\mathrm{b}}$(\footnote{e-mail: herve.bergeron@u-psud.fr, 
 evaldo@cbpf.br,
gazeau@apc.univ-paris7.fr, ligia@cbpf.br} )\\
\emph{$^{\mathrm{a}}$
Univ Paris-Sud, ISMO, UMR 8214, 91405 Orsay, France} \\
\emph{   $^{\mathrm{b}}$ Centro Brasileiro de Pesquisas Fisicas   } \\
\emph{   $^{\mathrm{c}}$ Instituto Nacional de Ci\^encia e Tecnologia - Sistemas Complexos}\\
\emph{  Rua Xavier Sigaud 150, 22290-180 - Rio de Janeiro, RJ, Brazil  } \\
\emph{  $^{\mathrm{d}}$ APC, UMR 7164,}\\
\emph{ Univ Paris  Diderot, Sorbonne Paris Cit\'e,  }  
\emph{75205 Paris, France} 
}

\maketitle
{\abstract{ 
 Asymptotic behavior (with respect to the number of trials) of symmetric generalizations of  binomial distributions and their related entropies are studied through three examples. The first one derives from the $\texttt{q}$-exponential  as a generating function. The second one involves the modified Abel polynomials, and the third one involves  Hermite polynomials. The former and the latter have extensive Boltzmann-Gibbs whereas the second one (Abel) has extensive R\'enyi entropy. A probabilistic model is presented for this exceptional case. 
}}


\section{Introduction}
\label{intro}

The content of  our previous papers \cite{curadoetal2011,bergeronetal2012,bergeronetal2013A,bergeronetal2013B}  was devoted  to a comprehensive study of discrete distributions generalizing the familiar  Bernoulli-like (or binomial-like) distributions. The generalization consists in substituting the ordinary integers on which is based the binomial distribution with arbitrary sequences of positive numbers. They can be symmetrical or asymmetrical. The study concerned the positiveness of those formal distributions in order to view them as having a real probabilistic content. We have given many examples, which run from Delone sequences, $q$-sequences, sequences based on family of polynomials (modified Abel, Hermite...). A key point of our works was to display manageable generating functions. The existence of such functions allows to  easily control positiveness and  makes a series of computations easier. Hence, we have shown in the above references a palette of interesting properties. Nevertheless, except in a few cases, we did not give illustrating models of these new probabilities distributions, and we did not explore systematically their asymptotic behaviors, their associated entropies (Shannon or Boltzmann-Gibbs, Tsallis, R\'enyi ...),  and related questions like extensiveness. 

The aim of  the present article is to examine comprehensively asymptotic  behaviors and associated entropies in the three cases concerning symmetric deformations of the binomial distribution previously presented in \cite{bergeronetal2013A,bergeronetal2013B} and having a sound probabilistic content. Our interest is particularly concerned with the extensivity, asymptotic or not, of these three entropies.  We recall  that an entropy is extensive (resp. asymptotically extensive) if it is proportional (resp. asymptotically proportional) to the number $n$ of events (resp. at large $n$). The three  probability distributions mentioned above are denoted in this paper by
\begin{equation}
\label{probdistG}
\mathcal{P}= \left( \mathfrak{p}^{(n)}_1, \mathfrak{p}^{(n)}_2, \dotsc \mathfrak{p}^{(n)}_n\right)\, . 
\end{equation}
Due to symmetry, the multiplicity of states is the same as for the binomial distribution. In our evaluations of entropies,  we adopt a ``microscopic" point of view by ignoring  the multiplicity. 

The first entropy is the Boltzmann-Gibbs (BG) or Shannon \cite{shannon48,shannon49} entropy.
\begin{equation}
\label{shannonBG}
S_{\mathrm{BG}}= - \sum_{k=0}^n \mathfrak{p}^{(n)}_k\,\log\frac{\mathfrak{p}^{(n)}_k}{\binom{n}{k}}\, .
\end{equation}
The second one is  the Tsallis entropy $S_q$ \cite{tsallis88}, which is  a deformation of \eqref{shannonBG}, $S_q \to S_{\mathrm{BG}}$ as $q\to 1$,
\begin{equation}
\label{tsallisq}
S_{q}= \frac{1}{q-1}\left\lbrack 1- \sum_{k=0}^n \binom{n}{k}\left(\frac{\mathfrak{p}^{(n)}_k}{\binom{n}{k}}\right)^q\right\rbrack\, .
\end{equation} 
The third one is the   R\'enyi entropy $S_{\mathrm{Re};q}$ \cite{renyi60}, which is also a deformation of \eqref{shannonBG} $ S_{\mathrm{Re};q} \to S_{\mathrm{BG}}$ as $q\to 1$,
  \begin{equation}
\label{renyiq}
S_{\mathrm{Re};q}= \frac{1}{1-q} \log\left\lbrack\sum_{k=0}^n \binom{n}{k} \left(\frac{\mathfrak{p}^{(n)}_k}{\binom{n}{k}}\right)^q\right\rbrack\, .
\end{equation}

The organization  of the paper is the following. In Section \ref{sec:genset}  the necessary background issued from \cite{bergeronetal2013B} is rewieved.  The first  case, examined in Section \ref{sec:qexp}, has the so-called $\texttt{q}$-exponential as a generating function. It gives rise to a nice probabilistic interpretation (e.g. Polya urns) and to  an extensive Boltzmann-Gibbs entropy, as was already mentioned in \cite{bergeronetal2013B}. We show that the R\'enyi entropy is also extensive for this distribution.  The second case is related to modified Abel polynomials and has the exponential of the Lambert function as a generating function. This example forms the  content of Section \ref{sec:abelpol} and yields an unexpected nontrivial result. Indeed, the entropy which is asymptotically extensive in this case is not Boltzmann-Gibbs, nor the Tsallis $q$-entropy for any $q$, but instead the R\'enyi one, and its asymptotic behavior does not depend on the R\'enyi parameter. Due to the importance of this result, we present in the same section a probabilistic model based on counting of words made with letters picked in several alphabets. This model is quite elaborate.  Section \ref{sec:hermpol} is devoted to our third example, involving Hermite polynomials. With this case, we return  to  the standard situation for which both  Boltzmann-Gibbs and R\'enyi are extensive. Following our conclusions and comments in Section \ref{sec:conclus} are the first  appendix where we present an historical survey of the concept of entropy, and the second one where we give the necessary technical details.

\section{Symmetric deformations of binomial distributions}
\label{sec:genset}
We remind in this section notations and main results of \cite{bergeronetal2013B}. 

Let  $\mathcal{X} = (x_n)_{n\in \N}$ be a sequence of positive numbers $x_n$ for $n>0$ and $x_0 = 0$.  
The ``factorial" of $x_n$  is defined as $x_n! = x_1\, x_2\, \cdots x_n\, ,\quad x_0! \deq 1\, ,$ and  from it we build the binomial coefficient
\begin{equation*}
\binom{x_n}{x_k}  := \frac{x_n!}{x_{n-k}! x_k!}\, . 
\end{equation*}
We now associate to  $\mathcal{X}$ the formal distribution
\begin{equation}
\label{symdist1}
\mathfrak{p}_k^{(n)}(\eta)=\binom{x_n}{x_k} \, q_k(\eta) q_{n-k}(1-\eta)\,,
\end{equation}
where the $q_k(\eta)$ are polynomials of  degree $k$
and the  $\mathfrak{p}_k^{(n)}(\eta)$ are constrained by the \emph{normalization} condition 
\begin{equation}
\label{normdist}
\forall n \in \mathbb{N}, \quad \forall \eta \in [0,1], \quad \sum_{k=0}^n \mathfrak{p}_k^{(n)}(\eta)=1,
\end{equation}
and by the 
\emph{non-negativeness} condition 
\begin{equation}
\label{nonnegcond}
\forall n,k \in \mathbb{N}, \quad \forall \eta \in [0,1], \quad \mathfrak{p}_k^{(n)}(\eta) \ge 0\, .
\end{equation}
The normalization  implies
 \begin{equation*}
 \forall \eta \in [0,1], \quad \mathfrak{p}_0^{(0)}(\eta)=q_0(\eta) q_0(1-\eta)=1 \Rightarrow  q_0(\eta)=\pm 1
\end{equation*}
From now on we keep the choice $q_0(\eta)=1$. This  implies 
\begin{equation*}
\forall n \in \mathbb{N}, \quad \forall \eta \in [0,1], \quad \mathfrak{p}_0^{(n)}(\eta)=q_n(1-\eta).
\end{equation*}
Therefore the non-negativeness condition  is equivalent to the non-negativeness of the polynomials $q_n$ on the interval $[0,1]$.
 The quantity $\mathfrak{p}_k^{(n)}(\eta)$  can be interpreted as the probability of having $k$ wins and $n-k$ losses in a sequence of \emph{correlated} $n$ trials.  Besides, as we recover the  invariance under  $k \to n-k$ and $\eta \to 1-\eta$ of the binomial distribution, no bias in  the case $\eta = 1/2$ can exist favoring either win or loss.
 
We now  associate to the sequence $\mathcal{X}$ an  ``exponential"  defined as the entire series
\begin{equation}
\label{formexp}
\mathcal{N}(t) = \sum_{n=0}^{\infty} \frac{ t^n}{x_n!}\equiv \sum_{n=0}^\infty a_n t^n\, ,\quad x_n=a_{n-1}/a_n\, ,  
\end{equation}
which is supposed  to have a non-vanishing radius of convergence. 
Hence $\mathcal{N}(\mathrm{t}) $ is an element of  $\Sigma$ defined as
 the set of entire series $\sum_{n=0}^\infty a_n t^n$ possessing a non-vanishing radius of convergence and verifying $a_0=1$ and $\forall n \ge1, \, a_n>0$.

 Starting from  $\mathcal{N}(t) \in \Sigma$ and $\eta \in [0,1]$, we consider the series $\mathcal{N}(t)^\eta$. It is easy to prove from $\mathcal{N}(t) = \mathcal{N}(t)^\eta\, \mathcal{N}(t)^{1-\eta}$ that  it is a  generating function for polynomials $q_n$ obeying \eqref{symdist1}-\eqref{normdist}:
\begin{equation}
\label{gensym}
\forall \eta \in [0,1], \quad \tilde G_{\mathcal{N},\eta}(t):= \mathcal{N}(t)^\eta=\sum_{n=0}^\infty \frac{q_n(\eta)}{x_n!} t^n.
\end{equation}
More precisely, the  polynomials $q_n$ issued from \eqref{gensym} have the following properties:
\begin{itemize}
\item[(a)] $q_0(\eta)=1$, $q_1(\eta)=\eta$ and more generally 
\begin{equation}
\label{eq:qnrecurrence}
\begin{split}
\forall n \in \mathbb{N}\, ,  \, \forall \eta \in [0,1]\, , \ q_{n+1}(\eta)&= \eta \, \dfrac{x_{n+1}}{n+1} \times \\   &\times \sum_{k=0}^n \left(\begin{array}{c}x_n \\ x_k \end{array}\right) \dfrac{n-k+1}{x_{n-k+1}}\,  q_k(\eta-1)\,.
\end{split}
\end{equation}
\item[(b)] The $q_n$'s are polynomials of degree $n$ obeying
\begin{equation*}
\forall n \in \mathbb{N}\, , \ q_n(1)=1, \ {\rm and } \ \forall n\ne 0\, , \, q_n(0)=0\, ,
\end{equation*}
 and they fulfill the normalization condition.
\item[(c)] The $q_n$'s fulfill the functional relation
\begin{equation}
\label{eq:fctrelation}
\forall z_1, z_2 \in \mathbb{C}, \, \forall n \in \mathbb{N}, \, \sum_{k=0}^n \left(\begin{array}{c}x_n \\ x_k \end{array}\right) q_k(z_1) q_{n-k}(z_2)=q_n(z_1+z_2)\, .
\end{equation}
\end{itemize}
We note that these polynomials, suitably normalized, are of binomial type.

Since $\log (\mathcal{N}(t))$ is analytical around $t=0$,  $\mathcal{N}(t)^\eta=\exp( \eta \log (\mathcal{N}(t)))$ possesses a convergent series expansion around $t=0$ (for all $\eta \in \mathbb{C}$). 

Since we already know that $q_0(\eta)=1$ and $\forall\;n\ne 0, q_n(0)=0$, the non-negativeness condition is equivalent to specify that for any $\eta \in ]0,1], \, q_n(\eta)>0$ and then the function $t \mapsto \mathcal{N}(t)^\eta$ belongs to $\Sigma$. Defining $\Sigma_0$ as the set of entire series $f(z)=\sum_{n=0}^\infty a_n z^n$ possessing a non-vanishing radius of convergence and verifying the conditions $a_0=0$, $a_1 >0$ and $\forall n \ge 2, \, a_n \ge 0$, 
it was proved in \cite{bergeronetal2013B} that 
\begin{equation}
\label{sigma+}
\Sigma_+:=\left\{ \mathcal{N} \in \Sigma \, | \,  \forall \eta \in [0,1), \,  \forall n \geq 0, \, q_n(\eta) > 0 \, \right\}=\{ e^F \, | \, F \in \Sigma_0\}
\end{equation}
is the set of deformed exponentials such that the generating functions $G_{\mathcal{N},\eta}(t)$ solve the non-negativeness problem.

\section{Symmetric distribution from``$\texttt{q}$-exponential''}
\label{sec:qexp}
\subsection{The probabilty distribution}
We consider here the following family of functions belonging to $\Sigma_+$: 
\begin{equation}
\label{genex1}
\NN (t) = \left( 1- \frac{t}{\alpha}\right)^{-\alpha}\, , \quad \alpha > 0\, , 
\end{equation}
that are $\texttt{q}$-exponentials in the sense that  $e_\texttt{q}(x) = [1-(1-\texttt{q})  x]^{(1/(1-\texttt{q}))}$, where the 
parameter 
$\texttt{q}=1+1/\alpha$ with the notations in  \cite{tsallisBJP09}. 
We first note that if $\alpha \to \infty$ then $\NN(t) \to e^t$, i.e. we return to the ordinary binomial case. The corresponding sequence is bounded by $\alpha$ and  given by
\begin{equation}
\label{xnex1}
x_n = \frac{n\alpha}{n + \alpha -1}\, , \quad \lim_{n\to \infty}x_n = \alpha \, . 
\end{equation}
For the factorial we have:
\begin{equation}
\label{factxn1}
x_n! = \alpha^n \frac{\Gamma(\alpha) n!}{\Gamma(n+\alpha)}=  \frac{ \alpha^n n!}{(\alpha)_n} = \frac{\alpha^n}{\binom{n+\alpha -1}{n}}\, ,
\end{equation}
where $(z)_n = \Gamma(z+n)/\Gamma(z)$ is the Pochhammer symbol.
The corresponding polynomials are   given by
\begin{equation}
\label{xnex1}
q_n(\eta) =  \frac{\Gamma (\alpha)}{\Gamma (n+ \alpha)}\, \frac{\Gamma (n+ \alpha \eta)}{\Gamma (\alpha \eta)}=  \frac{ (\alpha\eta)_n}{ (\alpha)_n}\, , 
\end{equation}
and satisfy the recurrence relation
\begin{equation}
\label{recrelex1}
q_n(\eta) =  \frac{ n + \alpha \eta -1}{n + \alpha -1}\, q_{n-1}(\eta)\, , \quad \mbox{with}\ q_0(\eta) = 1\, . 
\end{equation}
In particular $q_1(\eta) = \eta$. 
The  distribution $\mathfrak{p}_k^{(n)}(\eta)$ defined by these polynomials is given by
\begin{align}
\label{genbinsymA}
\mathfrak{p}_k^{(n)}(\eta)= &\binom{n}{k}\, \, \frac{\Gamma(\alpha)}{\Gamma(\eta\alpha)\Gamma((1-\eta)\alpha)}\, \frac{\Gamma(\eta\alpha +k)\Gamma((1-\eta)\alpha +n-k)}{\Gamma(\alpha +n)}\\
\label{genbinsymB} =&\frac{\displaystyle\binom{\eta \alpha + k-1}{k}\, \binom{(1-\eta) \alpha + n-k-1}{n-k}}{\displaystyle\binom{ \alpha +n-1}{n}}\, . 
\end{align}
This is precisely the P\'olya distribution \cite{johnson92}, also called  ``Markov-P\'olya" or ``inverse hypergeometric" and more. It was considered by P\'olya (1923) in the following urn scheme \cite{polya23}. From a set of $b$ black balls  and $r$ red balls  contained in an urn one extracts one ball and return it to the urn together with $c$ balls of the same color. The probability to have in the urn $k$ black balls after the $n$-th trial is given by the ratio \eqref{genbinsymB} with 
\begin{equation}
\label{polyaurn}
\eta= \frac{b}{b+r}\, , \qquad \alpha= \frac{b+r}{c}\, ,
\end{equation}
which holds for rational parameters $\eta$ and $\alpha$. In this notation, the distribution \eqref{genbinsymA} reads, in terms of Pochammer symbol,
\begin{equation}
\label{polya}
\mathfrak{p}_k^{(n)}(b,c,r)= \binom{n}{k}\, \frac{\displaystyle\left(\frac{b}{c}\right)_{k}\left(\frac{r}{c}\right)_{n-k}}{\displaystyle\left(\frac{b +r}{c}\right)_{n}} \, . 
\end{equation}

We notice that if we take the medium value  $\eta  = 1/2$ and redefine the parameters according to  $\alpha \rightarrow 2 \nu $, $n \rightarrow N$ and $k \rightarrow n$ in the distribution given by Eq. \eqref{genbinsymA} we recover the distribution $r_n^N$ 
studied in reference \cite{haneletal09}, see Eqs. (4) and (10) therein,  within the framework of the Laplace-de Finetti representation.

\subsection{Asymptotic behavior at large $n$}

Let us now study the asymptotic behavior of \eqref{genbinsymA} at large $n$. The  probability distribution is given by:
\begin{align*}
\mathfrak{p}_k^{(n)}(\eta)&= \binom{n}{k}\, \frac{\Gamma(\alpha)}{\Gamma(\eta\alpha)\Gamma((1-\eta)\alpha)}\, \frac{\Gamma(\eta\alpha +k)\Gamma((1-\eta)\alpha +n-k)}{\Gamma(\alpha +n)}\\ &= \binom{n}{k}\,\frac{B(\eta \alpha + k, (1-\eta)\alpha + n-k)}{B(\eta\alpha, (1-\eta)\alpha)}\, ,
\end{align*}
where $0\leq \eta\leq 1$, $\alpha>0$, and $B(p,q) = \Gamma(p)\Gamma(q)/\Gamma(q+p)$ is the beta function .
We put $k= nx$, with $0<x<1$. Using the Stirling formula, $n!\sim\sqrt{2\pi} \,e^{-n}\,n^{n+1/2}$ or $\Gamma(z)\sim \sqrt{2\pi} \,e^{(z-1/2)\log z-z}$, we find
\begin{equation*}
B(\eta \alpha + k, (1-\eta)\alpha + n-k) \sim \sqrt{\frac{2\pi}{n}}\,
 x^{\eta\alpha-1/2}\,(1-x)^{(1-\eta)\alpha-1/2}\, e^{-n\mathcal{C}(x)}\,, 
 \end{equation*}
 where we introduced  
\begin{equation}
\label{fctcx}
\begin{split}
 \mathcal{C}(x) &:=- x\log x - (1-x) \log(1-x)\,  ,\\ \text{with} \quad &\mathcal{C}^{\prime}(x) = -\log\frac{x}{1-x}\, , \quad  \mathcal{C}^{\prime\prime}(x)= -\frac{1}{x(1-x)}\, . 
\end{split}
\end{equation}
For $x\in (0,1)$ this function is nonnegative, concave  and symmetric with respect to its maximum value $\log 2$ at $x=1/2$. In  fact, $ \mathcal{C}(x)$ is the basic BG (or Shannon) entropy in  the case of two possibilities with probabilities $x$ and $1-x$, and it  appears in many places in the paper.    
 
The  asymptotic behavior of the binomial coefficient at large $n$ is (see \eqref{asymptbin})
\begin{equation*}
\nonumber \binom{n}{k=nx} \sim \frac{1}{\sqrt{2\pi nx(1-x)}}\, e^{n\mathcal{C}(x))} \, . 
\end{equation*}
Therefore, the limit distribution we find is the following:
\begin{equation}
\label{qexplimA}
\mathfrak{p}_{k= nx}^{(n)}(\eta) \sim \frac{1}{n}\, \frac{1}{B(\eta\alpha,(1-\eta)\alpha)}\, x^{\eta\alpha-1}\,(1-x)^{(1-\eta)\alpha-1}\,. 
\end{equation}
We easily check that the probabilistic normalisation $\sum_{k=0}^n\mathfrak{p}_{k= nx}^{(n)}=1$ remains valid at the limit $n\to \infty$. Indeed, replacing  the sum $\sum_{k=0}^n $ by the integral $\int_0^1 n\ud x$ leads to
\begin{equation}
\label{binsumint}
\sum_{k=0}^n \mathfrak{p}_{k= nx}^{(n)} \sim \frac{1}{B(\eta\alpha,(1-\eta)\alpha)}\int_{0}^1 x^{\eta\alpha-1}\,(1-x)^{(1-\eta)\alpha-1}\, \ud x= 1\,. 
\end{equation}

Moreover, our asymptotic formula \eqref{qexplimA}, in the case  $\eta = 1/2$ and  after centering on the origin, becomes proportional to a $Q$-Gaussian \cite{prato99} with $Q= (\alpha -4)/(\alpha -2)$. This result was recently obtained numerically by Ruiz and Tsallis \cite{ruiz14}.

\subsection{Boltzmann-Gibbs entropy}

\begin{figure}[htb!]
\begin{center}
\includegraphics[width=2.6in]{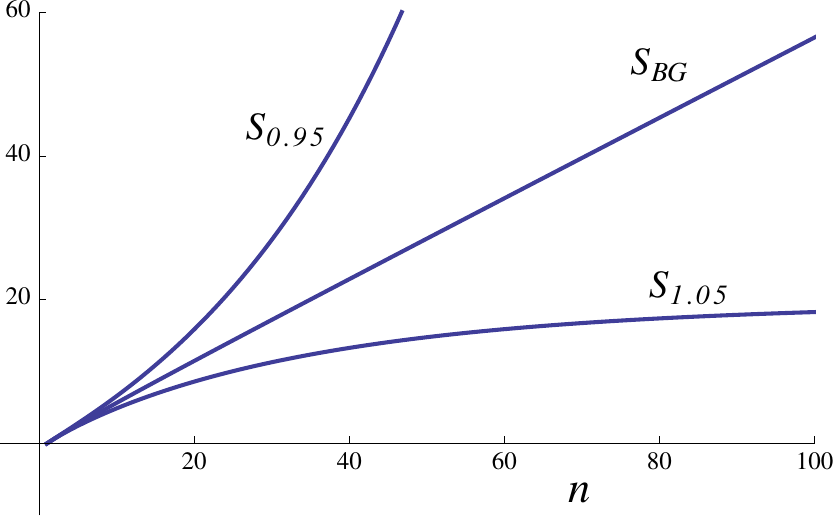}
\caption{$n$-dependence of entropies for the  ``$q$-exponential'' case 
($\mathcal{N}(t) = (1-t/\alpha)^{-\alpha}$).  The Boltzmann-Gibbs (BG) and Tsallis ($S_q$) entropies of 
the distribution are shown  for $\eta = 1/2$ and $\alpha = 3$. Upper curve is for $q=0.95$. Bottom curve is for $q=1.05$.
}
\label{entropy}
\end{center}
\end{figure}

We take a definition of the BG entropy for the distribution (\ref{genbinsymA}) which does not take into account the multiplicity of states, because  as we have already mentioned in the introduction, we are adopting a microscopic point of view. Consequently we replace the random variable $-\log\mathfrak{p}^{(n)}_k$ by $-\log\left(\mathfrak{p}^{(n)}_k/\binom{n}{k}\right)$:
\begin{equation}
\label{BGdefqexp}
S_{\mathrm{BG}}= - \sum_{k=0}^n \mathfrak{p}^{(n)}_k\,\log\frac{\mathfrak{p}^{(n)}_k}{\binom{n}{k}}\, .
\end{equation}
The division  of the probability by the binomial coefficient in each logarithm  in (\ref{BGdefqexp}) means a counting of the degeneracy. As a preliminary numerical exploration, its extensive property is shown in Figure \ref{entropy} where it is  compared with  the Tsallis entropy  $$S^{(n)}_q \deq (1- \sum_{k=0}^{n}\binom{n}{k} \left(\frac{\mathfrak{p}^{(n)}_k}{\binom{n}{k}}\right)^q)/(q-1)\,.$$ 
Let us now establish the analytic formula for
$S_{\mathrm{BG}}$ in the asymptotic limit as $n\to \infty$.  In the present case the latter 
behaves  as
\begin{align*}
& S_{\mathrm{BG}}\underset{\mbox{at large} \, n}{\sim} -  \frac{1}{B(\eta\alpha,(1-\eta)\alpha)}\int_{0}^1 \ud x\,x^{\eta\alpha-1}\,(1-x)^{(1-\eta)\alpha-1}\times \\&\times 
\log\left\lbrack\frac{1}{B(\eta\alpha,(1-\eta)\alpha)} \sqrt{\frac{2\pi}{n}}\,
 x^{\eta\alpha-1/2}\,(1-x)^{(1-\eta)\alpha-1/2}\, e^{-n\mathcal{C}(x))} \right\rbrack\\
 &= nI_1 + \frac{1}{2}\log n + I_2\, , 
\end{align*}
with
\begin{equation*}
I_1= \frac{1}{B(\eta\alpha,(1-\eta)\alpha)}\int_{0}^1 \ud x\,x^{\eta\alpha-1}\,(1-x)^{(1-\eta)\alpha-1}\,\mathcal{C}(x)\, , 
\end{equation*}
\begin{align*}
I_2&= \log\left(\frac{B(\eta\alpha,(1-\eta)\alpha)}{\sqrt{2\pi}}\right) - \frac{1}{B(\eta\alpha,(1-\eta)\alpha)}\int_{0}^1 \ud x\,x^{\eta\alpha-1}\,(1-x)^{(1-\eta)\alpha-1}\times \\
&\times\left[\left(\eta\alpha -\frac{1}{2}\right)\log x + \left((1-\eta)\alpha -\frac{1}{2}\right)\log(1-x)\right]\, . 
\end{align*}
Since the  term in $n$ is dominant, the Boltzmann-Gibbs entropy is proved to be extensive in the present case. Let us calculate the integrals appearing in the above expressions. They are all of the type
\begin{align}
\label{betalog}
\nonumber LB(p,q)&:= \int_{0}^1 \ud x\,  x^{p-1} \,\log x \, (1-x)^{q-1} = \int_{0}^1 \ud x \, (1-x)^{p-1} \,\log (1-x)  \, x^{q-1}\\
&= \frac{\partial}{\partial p} B(p,q) = [\psi(p)-\psi(p+q)]\,B(p,q)\, , \quad \psi(t) = \frac{\Gamma^{'}(t)}{\Gamma(t)}\, . 
\end{align}
Finally, we find
\begin{align}
\label{BGqexpasymp}
\nonumber S_{\mathrm{BG}}&\sim n [\psi(\alpha +1) -(\eta\psi(\eta \alpha +1) + (1-\eta)\psi((1-\eta) \alpha +1)] + \frac{1}{2}\log n +
\\ \nonumber 
&+ \log\left(\frac{B(\eta\alpha,(1-\eta)\alpha)}{\sqrt{2\pi}} \right)\\&  + \alpha \psi(\alpha) -\left[\left(\eta \alpha - \frac{1}{2}\right)\psi(\eta\alpha)
+ \left((1-\eta) \alpha - \frac{1}{2}\right)\psi((1-\eta)\alpha)\right]\,. 
\end{align}

\subsection{Rényi Entropy}
We finally explore, for the present case,  the  R\'enyi entropy
 \begin{equation}
\label{Renyiqexp}
S_{\mathrm{Re};q}= \frac{1}{1-q} \log\left\lbrack\sum_{k=0}^n \binom{n}{k} \left(\frac{\mathfrak{p}^{(n)}_k}{\binom{n}{k}}\right)^q\right\rbrack\, ,
\end{equation}
which becomes $S_{\mathrm{BG}}$ as $q\to 1$. Using the asymptotic formula of Eqs.\,\eqref{asymptbin} and \eqref{qexplimA}, the approximation $\sum_{k=0}^n \sim \int_0^1 n \ud x$, and the Laplace formula (see \eqref{BGhermasymplap}), we obtain the asymptotic expression for $q < 1$ 
\begin{equation}
\label{asymptsumqexp}
\sum_{k=0}^n \binom{n}{k} \left(\frac{\mathfrak{p}^{(n)}_k}{\binom{n}{k}}\right)^q \sim \frac{1}{2 \sqrt{1-q}} \left\lbrack \frac{2^{5-2\alpha} \pi}{n B^2(\eta \alpha, (1-\eta)\alpha)} \right\rbrack^{q/2} e^{n (1-q) \log2} \,.
\end{equation}
By taking the logarithm of \eqref{asymptsumqexp}, we see that  the dominant term is 
\begin{equation}
\label{Renyiqexp}
S_{\mathrm{Re};q} \sim n \log 2 \,.
\end{equation}
and  the R\'{e}nyi entropy is obviously extensive.
A  point to be noticed is that this asymptotic behavior is independent of the R\'{e}nyi parameter $q$. Actually this remarkable feature is encountered in many distributions \cite{bergeronetal2014A}, including the next two cases considered in this paper.  We will give  a special attention to this fact in Section  \ref{sec:conclus}.

\section{Symmetric distribution from  modified Abel polynomials}
\label{sec:abelpol}
\subsection{Probability distribution}
We take here the specific generating  function $\NN (t)$ given by
\begin{equation}
\label{genex2}
\NN (t) =e^{-\alpha W(-t/\alpha)}\, , \quad \alpha > 0\, , 
\end{equation}
where $W$ is the Lambert function \cite{lambert}, i.e.  solving  the functional equation $W(t)e^{W(t)} = t$. 
We first note that if $\alpha \to \infty$ then $\NN(t) \to e^t$. The corresponding sequence is bounded by $\alpha/e$ and  given by
\begin{equation}
\label{xnex2}
x_n = \frac{n\alpha}{n + \alpha }\left(1- \frac{1}{n + \alpha}\right)^{n-2}\, , \quad \lim_{n\to \infty}x_n = \alpha/e \, . 
\end{equation}
We also note that $x_n \to n$ as $\alpha \to \infty$.  The corresponding factorial is
\begin{equation}
\label{factabel}
x_n!= n!\, \frac{\alpha^{n-1}}{(n+\alpha)^{n-1}}\, .
\end{equation}
The  polynomials $q_n$'s read as
\begin{equation}
\label{xnex2}
q_n(\eta) =  \eta \frac{\left(\eta + \frac{n}{\alpha}\right)^{n-1}}{\left(1 + \frac{n}{\alpha}\right)^{n-1}}\, . 
\end{equation}
We verify that  $q_0(\eta) = 1$ and $q_1(\eta) = \eta$. 
The polynomials above are a kind of modified Abel polynomials \cite{roman84} which look like 
\begin{equation}
\label{abel}
P_n(x) = x(x+na)^{n-1}\, , a\in \mathbb{Q}\, , 
\end{equation}
with the difference of the presence of a normalization factor in the denominator of \eqref{xnex2} and the relaxing of the rational condition. 

The corresponding probability distribution is found to be:
\begin{equation}
\label{abelprobdist}
\mathfrak{p}_k^{(n)}(\eta) = \binom{n}{k}\eta(1-\eta) \frac{(\eta + k/\alpha)^{k-1}(1-\eta + (n-k)/\alpha)^{n-k-1}}{(1+n/\alpha)^{n-1}}\,, 
\end{equation}
with $0\leq \eta\leq 1$.
\subsection{Regularization at the limit $n\to \infty$}
Putting $k= nx$ in \eqref{abelprobdist}, with $0<x<1$, and using the Stirling formula, we find the limit distribution
\begin{equation}
\label{limabelprob}
\mathfrak{p}_{nx}^{(n)}(\eta)\underset{\mbox{large $n$}}{\sim}\, \frac{\alpha \eta(1-\eta)}{\sqrt{ 2\pi}}(nx(1-x))^{-3/2}\,.
\end{equation}
The problem is that if one replaces  the discrete sum $\sum_{k=0}^n$ by the integral $\int_0^1 n\mathrm{d}x$ this expression leads to a divergent integral. However, there is a simple way to give it a finite value through a sort of principal value. First, let us consider the finite convergent integral  
\begin{equation}
\label{incbeta}
B_{\epsilon}(a,a):= \int_{\epsilon}^{1-\epsilon}x^{a-1}(1-x)^{a-1}\mathrm{d}x
\end{equation}
with  a small $\epsilon>0$ and arbitrary $a$. Due to the symmetry of the integrand under the interchange $x \to 1-x$, we have 
\begin{equation}
\label{incbeta1}
B_{\epsilon}(a,a)= 2\int_{\epsilon}^{\frac{1}{2}}x^{a-1}(1-x)^{a-1}\mathrm{d}x\,. 
\end{equation}
By expanding the binomial $(1-x)^{a-1}$ we easily find its expression in terms of Gauss hypergeometric function:
\begin{equation}
\label{incbeta2}
B_{\epsilon}(a,a)= \frac{2}{a}\left\lbrack 2^{-a} {}_{2}F_1\left(a,1-a;a+1; \frac{1}{2}\right) - \epsilon^{a} {}_{2}F_1\left(a,1-a;a+1; \epsilon\right)\right\rbrack\, . 
\end{equation}
We now consider our particular case $a= -1/2$. By using the formula \cite{magnus66}
\begin{equation}
\label{spechyper}
_{2}F_1\left(a,1-a;b; \frac{1}{2}\right) =\sqrt{\pi} 2^{1-b}\Gamma(b)\left\lbrack \Gamma\left(\frac{a+b}{2}\right)\Gamma\left(\frac{1+b-a}{2}\right)\right\rbrack^{-1}\, , 
\end{equation}
and from $1/\Gamma(0) = 0$, ${}_{2}F_1\left(a,1-a;a+1; \epsilon\right) \approx 1$ at small $\epsilon$, we eventually find 
\begin{equation}
\label{incbeta3}
B_{\epsilon}\left(-\frac{1}{2},-\frac{1}{2}\right)\sim  \frac{4}{\sqrt{\epsilon}}\, . 
\end{equation}
Now, from \eqref{limabelprob}, we have 
\begin{align}
\label{limabelprobeps}
\sum_{k=0}^n\mathfrak{p}_{nx}^{(n)}(\eta) &\underset{\mbox{large $n$}}{\sim}\, \frac{\alpha \eta(1-\eta)}{\sqrt{ 2\pi}}\,n^{-1/2}\,\lim_{\epsilon\to 0}\int_{\epsilon}^{1-\epsilon}x^{-\frac{3}{2}}(1-x)^{-\frac{3}{2}}\mathrm{d}x \\
&\underset{\mbox{large $n$}}{\sim}\,\lim_{\epsilon\to 0} \frac{4\alpha \eta(1-\eta)}{\sqrt{ 2\pi}} \frac{1}{\sqrt{n \epsilon}}\,. 
\end{align}
It is then legitimate to put $\epsilon = A/n$, where the arbitrarily  constant $A$ is consistently chosen as $A= 8(\alpha \eta(1-\eta))^2/\pi$, in such a way that the original expression remains equal to 1. 

\subsection{Boltzmann-Gibbs Entropy}
We first examine the Boltzmann-Gibbs entropy for the limit distribution \eqref{limabelprob}. From \eqref{limabelprob} and \eqref{asymptbin} we find the asymptotic behavior of each term in the defining sum of the BG entropy:
\begin{align}
\label{asympAbelBGterm}
\nonumber \mathfrak{p}^{(n)}_k\,\log\frac{\mathfrak{p}^{(n)}_k}{\binom{n}{k}} &\sim \frac{\alpha \eta(1-\eta)}{\sqrt{2\pi}}\,(nx(1-x))^{-3/2}\times\\
&\times \log\left\lbrack \alpha \eta(1-\eta) (nx(1-x))^{-1}e^{-n\mathcal{C}(x)}\right\rbrack\, . 
\end{align}

\begin{figure}[hbt!]
\begin{center}
\includegraphics[width=3in]{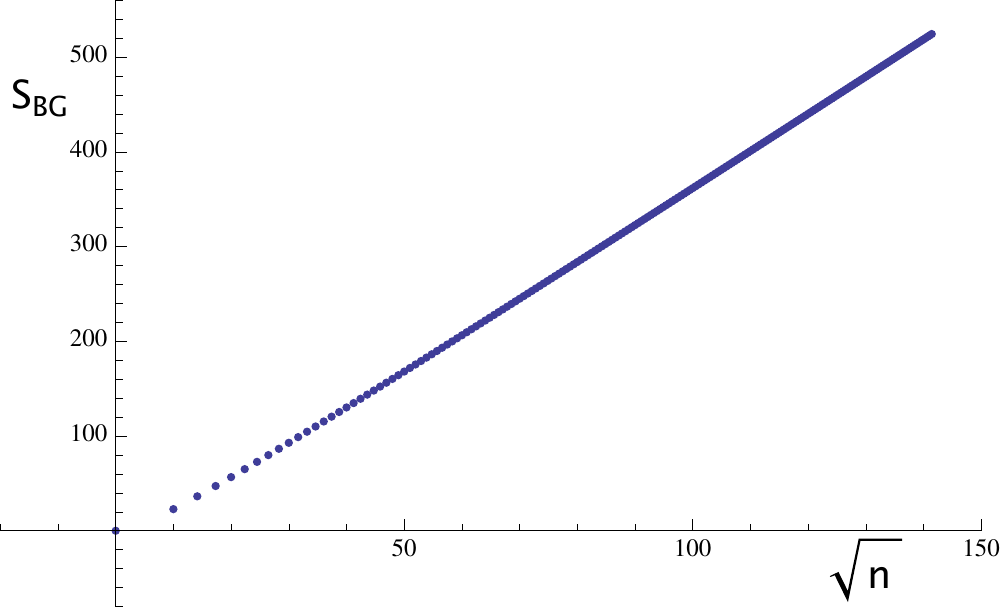}
\caption{Numerical behavior of the Boltzmann-Gibbs entropy \eqref{BGAbel1} versus $\sqrt{n}$ for the symmetric distribution from modified Abel polynomials, with $\eta =0.8$, $\alpha = 5$ and $n$ up to 20 000. }
\label{BGAbel}
\end{center}
\end{figure}

After  replacing  the sum $\sum_{k=0}^n $ by the integral $\int_0^1 n\ud x$  in 
\begin{equation}
\label{BGAbel1}
S_{\mathrm{BG}}= - \sum_{k=0}^n \mathfrak{p}^{(n)}_k\,\log\frac{\mathfrak{p}^{(n)}_k}{\binom{n}{k}}\, ,
\end{equation}
the BG entropy behaves   as the sum of three terms
\begin{equation}
\label{BGAbelRi}
 S_{\mathrm{BG}}\underset{\mbox{large $n$}}{\sim}\, \frac{1}{\sqrt{n}}  \left(\lim_{\epsilon_1 \to 0} R_{1;\epsilon_1} +  \lim_{\epsilon_2 \to 0}R_{2;\epsilon_2}\right) + \sqrt{n} R_3\, ,   
\end{equation}
where
\begin{align*}
\label{}
 R_{1;\epsilon_1}  &= -\frac{\alpha \eta(1-\eta)}{\sqrt{2\pi}} \log\left(\frac{\alpha \eta(1-\eta)}{n}\right) B_{\epsilon_1}\left(-\frac{1}{2},-\frac{1}{2}\right) \\
  R_{2;\epsilon_2}  &= 2\frac{\alpha \eta(1-\eta)}{\sqrt{2\pi}}LB_{\epsilon_2}\left(-\frac{1}{2},-\frac{1}{2} \right)\\
  R_3 &=   -2\frac{\alpha \eta(1-\eta)}{\sqrt{2\pi}} \int_0^1 x^{-1/2} \,(\log x)\, (1-x)^{-3/2}\, \ud x\, . 
\end{align*}
Here we have introduced the notation 
\begin{equation}
\label{ LBeps}
\begin{split}
LB_{\epsilon}\left(a,a\right) &:= \int_{\epsilon}^{1-\epsilon}x^{a-1}\,(\log{x})\,(1-x)^{a-1}\mathrm{d}x\\
&= \int_{\epsilon}^{1-\epsilon}x^{a-1}\,(\log{(1-x)})\,(1-x)^{a-1}\mathrm{d}x\, , 
\end{split}
\end{equation}
and made use of symmetries $x\mapsto 1-x$ in the integrals. From the general expression
\begin{align*}
LB_{\epsilon}\left(a,a\right) &= \frac{1}{2}\frac{\partial}{\partial a} B_{\epsilon}\left(a,a\right)  = -\frac{1}{2a} B_{\epsilon}\left(a,a\right) - \frac{\sqrt{\pi}}{a} 2^{1-2a}\log 2 \frac{\Gamma(a+1)}{\Gamma\left(a + \frac{1}{2}\right)} + \\
&+ \frac{\sqrt{\pi}}{a}2^{-2a}\,\frac{\Gamma(a+1)}{\Gamma\left(a + \frac{1}{2}\right)}\left[\psi(a+1)- \psi\left( a + \frac{1}{2}\right)\right] +\\
&- \frac{1}{a} \log\epsilon\,\epsilon^a\,{}_2F_1(a,1-a;a+1;\epsilon) - \frac{1}{a} \,\epsilon^a\,O(\epsilon)\, ,
\end{align*}
we find
\begin{equation*}
LB_{\epsilon}\left(-\frac{1}{2},-\frac{1}{2}\right) \sim \frac{4}{\sqrt{\epsilon}} +  \frac{2 \log \epsilon}{\sqrt{\epsilon}} \sim \frac{2 \log \epsilon}{\sqrt{\epsilon}}\, . 
\end{equation*}
In \eqref{BGAbelRi} choosing  $\epsilon_1 \propto 1/n$ and $\epsilon_2\, (\log\epsilon_2)^2 \propto 1/n$  we  get rid of both the integral divergences in $R_{1;\epsilon} $  and $R_{2;\epsilon}$ respectively.  The value of $R_3$ is easily found from \cite{gradry07}:
\begin{equation}
\label{LB3}
LI_p:= \int_{0}^{1}x^{p-1}\,(\log x)\, (1-x)^{-p-1}\, \ud x = -\frac{\pi}{p}\, \csc p\pi\, , \quad 0<p<1\,. 
\end{equation}
In the present case, $p= 1/2$ and so $LI_p = -2\pi$. 
Hence, we see that the dominant term in \eqref{BGAbelRi} is 
\begin{equation}
\label{BGAbeldom}
S_{\mathrm{BG}}\sim 2\sqrt{2\pi}\,\alpha\, \eta\,(1-\eta)\, \sqrt{n}\, . 
\end{equation}
Hence, we conclude that the Boltzmann-Gibbs entropy is not extensive for this type of deformation of the binomial distribution and with the chosen regularization of integrals. It behaves as 
$\sqrt{n}$, as is also shown in Figure \ref{BGAbel} obtained from numerical computations of \eqref{BGAbel1}.

\subsection{Tsallis Entropy}
Let us now explore, for the present case, the Tsallis entropy defined as
\begin{equation}
\label{TsallisAbel1}
S_{q}= \frac{1}{q-1}\left\lbrack 1- \sum_{k=0}^n \binom{n}{k}\left(\frac{\mathfrak{p}^{(n)}_k}{\binom{n}{k}}\right)^q\right\rbrack\, ,
\end{equation}
which becomes  $S_{\mathrm{BG}}$ as $q\to 1$. 
We first estimate the general term in the sum:
\begin{align}
\label{termsumabel}
 \binom{n}{nx} \left(\frac{\mathfrak{p}^{(n)}_{nx}}{\binom{n}{nx}}\right)^q&\sim \frac{(\alpha \eta(1-\eta))^q}{\sqrt{2\pi}}\,n^{-q -1/2}\,  (x(1-x))^{-q-1/2}\, e^{-n(q-1)\mathcal{C}(x)}\, .  
\end{align}
 We now replace  the sum $\sum_{k=0}^n $ by the integral $\int_0^1 n\ud x$ and obtain
\begin{equation}
\label{apsumabel}
\sum_{k=0}^n\binom{n}{k} \left(\frac{\mathfrak{p}^{(n)}_k}{\binom{n}{k}}\right)^q \sim \frac{(\alpha \eta(1-\eta))^q}{\sqrt{2\pi}}\,n^{1/2-q}\,  \int_0^1 (x(1-x))^{-q-1/2}\, e^{-n(q-1)\mathcal{C}(x)} \, \ud x\, , 
\end{equation}
and this would impose a convergence condition $q< 1/2$ if we were not in the very large $n$ regime. With the properties \eqref{fctcx} of $\mathcal{C}(x)$,  the use of the Laplace approximation method with the condition $q<1$ yields 
\begin{align}
\label{qabelsumasymp}
\sum_{k=0}^n\binom{n}{k} \left(\frac{\mathfrak{p}^{(n)}_k}{\binom{n}{k}}\right)^q&\sim \frac{1}{\sqrt{\vert q-1\vert}}\,2^{2q}(\alpha \eta(1-\eta))^q\, n^{-q}\, e^{n(1-q)\log 2}\, ,  
\end{align}
and  the Tsallis entropy becomes
\begin{equation}
\label{TsallisAbel2}
S_{q} \sim \frac{1}{(q-1) \sqrt{\vert q-1\vert}}\left\lbrack \sqrt{\vert q-1\vert}- 2^{2q}(\alpha \eta(1-\eta))^q\, n^{-q}\, e^{n(1-q)\log 2}\right\rbrack \, . 
\end{equation}
We see that the Tsallis entropy is not extensive for any value of $q< 1$. However, we should be aware that our derivation  prevents us  to consider the asymptotic form of  $S_{\mathrm{BG}}$ in \eqref{BGAbeldom} as the limit at $q\to 1$ of \eqref{TsallisAbel2}, since the Laplace approximation method in \eqref{apsumabel} loses its validity for $q=1$. 

\subsection{Rényi Entropy}
We finally explore the  R\'enyi entropy
 \begin{equation}
\label{RenyiAbel1}
S_{\mathrm{Re};q}= \frac{1}{1-q} \log\left\lbrack\sum_{k=0}^n \binom{n}{k} \left(\frac{\mathfrak{p}^{(n)}_k}{\binom{n}{k}}\right)^q\right\rbrack\, .
\end{equation}
From  \eqref{qabelsumasymp} we derive immediately
\begin{align}
\label{logtermsumabel}
\nonumber \log \left\lbrack\sum_{k=0}^n\binom{n}{nx} \left(\frac{\mathfrak{p}^{(n)}_{nx}}{\binom{n}{nx}}\right)^q\right\rbrack&\sim \log\left(\frac{1}{\sqrt{\vert q-1\vert}}\,2^{2q}(\alpha \eta(1-\eta))^q\right) + \\ &-q\log n + n(1-q)\log 2 \, . 
  \end{align}
Therefore, the R\'enyi entropy is extensive for $q<1$:
\begin{equation}
\label{RenyiAbel1}
S_{\mathrm{Re};q} \sim  n \, \log 2 \, .
\end{equation}
We recover the asymptotic $q$-independence already noticed in the case of the previous example. 

\subsection{Probabilistic interpretation}
Choosing the parameters $\alpha$ and $\eta$ in the expression \eqref{abelprobdist} as
\begin{equation}
\alpha = \frac{p+q}{c} \quad \text{and} \quad \eta = \frac{p}{p+q} \,,
\end{equation}
where $p$, $q$ and $c$ are three positive integers, we obtain
\begin{equation}
\label{proba}
\mathfrak{p}_k^{(n)} = \binom{n}{k} \frac{p (p + k c)^{k-1} q (q + (n-k) c)^{n-k-1}}{(p+q) (p+q +n c)^{n-1}}\,. 
\end{equation}
>From the sum of probabilities, we deduce the finite expansion formula
\begin{equation}
\label{sum}
\sum_{k=0}^n  \binom{n}{k} p (p + k c)^{k-1} q (q + (n-k) c)^{n-k-1} = (p+q) (p+q +n c)^{n-1} \,.
\end{equation}
We now present counting interpretation of this expansion and its resulting urn model. We define  a finite set for which the numbers $ \binom{n}{k} p (p + k c)^{k-1} q (q + (n-k) c)^{n-k-1} $ for $k=0,1 \dots$ correspond to  counting of partitions. As our main interest is to present at least one sound probabilistic model, 
for the sake of simplicity we consider  the case $c=1$.

\subsubsection{The model} Let $\mathcal{A}(2n,p,q) = \mathcal{A}_{C\ell}(2n)\cup \mathcal{A}_{\ell}(p)\cup \mathcal{A}_{C}(q)$ be an alphabet of $2n+p+q$ letters viewed as the union of three sub-alphabets: 
\begin{itemize}
\item $\mathcal{A}_{\ell}(p)$, $p \ge 1$, is a set $\{ b_1, b_2, \dots , b_{p}\}$ of $p$ letters which are only lowercase, by convention $\mathcal{A}_{\ell}(0)=\emptyset$,

\item $ \mathcal{A}_{C}(q)$, $q \ge 1$, is a set $\{C_1, C_2, \dots, C_{q}\}$ of $q$ letters which  are solely capital, by convention $\mathcal{A}_{C}(0)=\emptyset$,

\item  The family $\{ \mathcal{A}_{C\ell}(2n) \}_{n=1}^\infty$ where $ \mathcal{A}_{C\ell}(2n)=\bigcup_{i=1}^{n} \{a_i, A_i\}$ made of $2n$ mixed letters, built from a possible infinite sequence of pairs 
$$(a_1,A_1), \dots (a_i,A_i), \dots$$
 Each pair $(a_i,A_i)$ is made from the same letter in both sizes (lowercase and capital), and the letters are assumed to be different in different pairs, independently of their size. The inclusion   $\mathcal{A}_{C\ell}(2n) \subset \mathcal{A}_{C\ell}(2m)$ holds for any $n \le m$. \\
In the following we introduce also the lowercase part of $ \mathcal{A}_{C\ell}(2n)$ as $\mathcal{A}_{C \ell} ^\ell (n)= \bigcup_{i=1}^n \{a_i\}$ and the capital part of $ \mathcal{A}_{C\ell}(2n)$ as $\mathcal{A}_{C \ell} ^C (n)= \bigcup_{i=1}^n \{A_i\}$.

\item All letters, independently of their size, are assumed to be different: in $\mathcal{A}(2n,p,q)$ we have $n+p$ different lowercase letters  and $n+q$ different capital letters.

\end{itemize}

We consider the set of words $\mathcal{W}_n$ with $n$ letters picked from $\mathcal{A}(2n,p,q)$, built as $\mathcal{W}_n =\bigcup_{k=0}^n \mathcal{W}^n_k$ where the subsets $\mathcal{W}^n_k$ contain the words with $n$ letters, $k$ of them being lowercase and $n-k$ capital. The words are built with the following rules.
\begin{enumerate}
\item[(i)] Different orderings of letters are assumed to give different words,
 \item[(ii)] In a word in $\mathcal{W}^n_k$, starting from the left, the first lowercase letter  encountered (if $k \ne 0$) belongs to $\mathcal{A}_{\ell}(p)$, and the first capital letter encountered (if $k \ne n$) belongs to $\mathcal{A}_{C}(q)$ .
  \item[(iii)] In a word in $\mathcal{W}^n_k$, all the lowercase letters ($k \ne 0$) belong to $\mathcal{A}_\ell(p) \cup \mathcal{A}_{C \ell}^\ell (k)$, and all the capital letters ($k \ne n$) belong to $\mathcal{A}_C(q) \cup \mathcal{A}_{C \ell}^C(n-k)$. 
\end{enumerate}

\bigskip

Now let us evaluate the number of words $\mathcal{N}^n_k$ in $\mathcal{W}^n_k$.
\begin{itemize}
\item If $k=0$ the words contain exactly $n$ capital letters. The first one (from the left) belongs to $\mathcal{A}_C(q)$ and the $n-1$ remaining ones belong to $\mathcal{A}_C(q) \cup \mathcal{A}_{C \ell}^C(n) $. This gives
\begin{equation}
\mathcal{N}^n_0 = q (q+n)^{n-1} \,.
\end{equation}
\item If $k=1$, the words contain a unique lowercase letter that belongs to $\mathcal{A}_\ell(p)$, and $n-1$ capital letters. The first capital letter belongs to $\mathcal{A}_C(q)$, the $n-2$ remaining (capital letters) belong to $\mathcal{A}_C(q) \cup \mathcal{A}_{C \ell}^C(n-1) $. Since there is $n=\binom{n}{1}$ ways to locate the lowercase letter in the word, we have
\begin{equation}
\mathcal{N}^n_1 = \binom{n}{1} p q (q +n-1)^{n-2}
\end{equation}

\item For $ 2 \le k \le n-2$, we first  choose the $k$ positions of the lowercase letters in the word, there are $\binom{n}{k}$ possibilities. The first lowercase letter belongs to $\mathcal{A}_\ell(p)$, the following $k-1$ ones belong to $\mathcal{A}_\ell(p) \cup \mathcal{A}_{C \ell}^\ell (k)$, then for each choice of the $k$ positions, we have $p (p+k)^{k-1}$ possibilities for the lowercase letters. For the capital letters, we obtain similarly $q(q+n-k)^{n-k-1}$ possibilities. We deduce
\begin{equation}
\label{countinghb}
\mathcal{N}^n_k = \binom{n}{k} p (p+k)^{k-1} q (q +n-k)^{n-k-1}
\end{equation}
\item The cases $k=n-1$ and $k=n$ are analyzed following the same rules,  leading to
\begin{equation}
\mathcal{N}^n_{n-1} = \binom{n}{n-1} p q (p+n-1)^{n-2} \quad \text{and} \quad \mathcal{N}^n_{n} =  p (p+n)^{n-1} \,.
\end{equation}
\end{itemize}
We conclude that the formula of Eq.\eqref{countinghb} is valid for $k=0,1 \dots, n-1,n$.
Using Eq.\eqref{sum} we deduce that the total number $\mathcal{N}_n$ of words of $\mathcal{W}_n$ is $\mathcal{N}_n=(p+q) (p+q+n)^{n-1}$ . \\

\begin{remark} The value of $\mathcal{N}_n$ can be easily understood. A generic word of $\mathcal{W}_n$ contains:
\begin{itemize}
\item One letter that belongs either to $\mathcal{A}_\ell(p)$ or to  $\mathcal{A}_C(q)$: this gives $p+q$ possibilities,
\item Each remaining letter is either lowercase belonging to $\mathcal{A}_\ell(p) \cup \mathcal{A}_{C \ell}^\ell (k)$, or capital belonging to $\mathcal{A}_C(q) \cup \mathcal{A}_{C \ell}^C(n-k)$ for some $k$. This gives $(p+k)+(q+n-k)=p+q+n$ possibilities for each $n-1$ letters.
\end{itemize}
Therefore  $\mathcal{N}_n=(p+q) (p+q+n)^{n-1}$ .\\
\end{remark}
\begin{conclusion}
The probabilities $\frak{p}^{(n)}_k$ of Eq.\eqref{proba} are  the probabilities to extract a word with $k$ lowercase letters after  a draw at random from the ``urn" $\mathcal{W}_n$.
\end{conclusion}
\begin{remark} 
Other interesting probabilities emerge  from this urn model. 
For example let us call $P(\{l_1,l_2,\dots \})$ the probability that a word of $\mathcal{W}_n$ contains at least one of the letters of the family $\{l_1,l_2,\dots \}$. We have the following results
\begin{equation}
\left\{\begin{array}{c}
P(\mathcal{A}_\ell(p)) = 1-  \frak{p}^{(n)}_0 \\
P(\mathcal{A}_C(q)) = 1-  \frak{p}^{(n)}_n \\
\forall k \ge 2\, , \ P(\mathcal{A}_\ell(p) \cup \mathcal{A}_{C \ell}^\ell(k)) = \sum_{i=k}^n  \frak{p}^{(n)}_i \\
P(\mathcal{A}_\ell(p) \cup \mathcal{A}_{C \ell}^\ell(1)) = \frac{1}{2} P(\mathcal{A}_\ell(p) \cup \mathcal{A}_{C \ell}^\ell(2)) 
\end{array}\right.
\end{equation}
\end{remark} 
\subsubsection{An example}

Let us illustrate the above counting with the manageable although not trivial case  $n=3$, $p=q=1$ and the alphabet 
$$\mathcal{A}= \{a,\,  b,\,c,\, d,\, A,\, B,\, C,\, D\}\equiv  \mathcal{A}_{C \ell}(6) \cup \mathcal{A}_\ell(1) \cup \mathcal{A}_C(1)  ,$$ 
$$\mathcal{A}_{C \ell}(2)= \{a,\,  A\}\, , \, \mathcal{A}_{C \ell}(4)= \{a,\,  A\} \cup \{b,\, B\}\, , \,  \mathcal{A}_{C \ell}(6)= \{a,\,  A\} \cup \{b,\, B\} \cup \{c,\, C \} \,,$$
$$\mathcal{A}^{\ell}_{C \ell}(1) = \{a\}\, , \, \mathcal{A}^{\ell}_{C \ell}(2) = \{a,\, b\}\, ,\,  \mathcal{A}^{\ell}_{C \ell}(3) = \{a,\, b,\, c\}\, , $$
$$\mathcal{A}^{C}_{C \ell}(1) = \{A\}\, , \, \mathcal{A}^{C}_{C \ell}(2) = \{A,\, B\}\, , \, \mathcal{A}^{C}_{C \ell}(3) = \{A,\, B,\, C\}\, , $$
$$\mathcal{A}_\ell(1)= \{  d\}\, , \quad \mathcal{A}_C(1)= \{  D \}\,. $$
The total number of possible words of $\mathcal{W}_3$ is $\mathcal{N}_3=50$. The set of  allowed words  with 3 letters built from the above rules is described as follows. 
\begin{itemize}
  \item The subset of words $\mathcal{W}_0^3$ is 
 \begin{equation*}
\begin{pmatrix}
  DAA   & DAB &DAC & DAD  \\
  DBA  & DBB & DBC & DBD\\
  DCA & DCB & DCC & DCD\\
  DDA & DDB &  DDC & DDD \\
      \end{pmatrix} \, , 
\end{equation*}
corresponding to $\mathcal{N}_0^3=16$ words.

 \item The subset of words $\mathcal{W}_1^3$ is
 \begin{equation*}
\begin{pmatrix}
   dDA   & dDB & dDD   \\
   DdA   & DdB & DdD \\
   DAd & DBd & DDd \\
         \end{pmatrix} \, . 
\end{equation*}

corresponding to  $\mathcal{N}_1^3=9$ words.

 \item The subset of words $\mathcal{W}_2^3$ is
 \begin{equation*}
\begin{pmatrix}
   dDa   & dDb & dDd   \\
   daD   & dbD & ddD \\
   Dda & Ddb & Ddd \\
         \end{pmatrix} \, . 
\end{equation*}
corresponding to  $\mathcal{N}_2^3=9$ words.

  \item The subset of words $\mathcal{W}_3^3$ is 
 \begin{equation*}
\begin{pmatrix}
  daa   & dab & dac & dad  \\
  dba  & dbb & dbc & dbd\\
  dca & dcb & dcc & dcd\\
  dda & ddb &  ddc & ddd \\
      \end{pmatrix} \, , 
\end{equation*}
corresponding to $\mathcal{N}_3^3=16$ words.
\end{itemize}
The total number of words is $2 \times 16+ 2 \times 9=50$. 
Finally, the probabilities $ \mathfrak{p}_k^{(3)}$
 corresponding to these 4 situations are given in Table \ref{exprobabel}.
\begin{table}[h]
  \centering 
  \caption{values of $\mathfrak{p}_k^{(n)}$ for $n=3$, $\alpha = 2= p+q$, $\eta=p/(p+q)=1/2$, $p=q=1$}\label{exprobabel}
  \begin{tabular}{|c|c|}
\hline
  $k$ & $\mathfrak{p}_k^{(n)}$   \\
  \hline
 0  & 8/25   \\
 \hline
 1  &  9/50\\
 \hline
   2  &   9/50 \\
   \hline
   3    &  8/25 \\  
\hline
\end{tabular}
\end{table}

\section{Symmetric distribution from Hermite polynomials}
\label{sec:hermpol}
Here the function $\NN (t)$ is chosen as 
\begin{equation}
\label{genex2}
\NN (t) =e^{t+ \frac{a}{2}t^2}\, , \quad 0< a <1\, . 
\end{equation}
The corresponding sequence ${x_n}$ has the following factorial form:
\begin{equation}
\label{xnfat2} 
x_n! = \left[  \frac{i^n \left(\frac{a}{2}\right)^{n/2}}{n!} H_n \left( 
\frac{-i}{\sqrt{2 a}}
\right) \right]^{-1} = \left\lbrack\sum_{m=0}^{\left\lfloor\frac{n}{2}\right\rfloor} \frac{(a/2)^m}{m!(n-2m)!} \right\rbrack^{-1}:= \frac{1}{\varphi_{n}(a)}\,.
\end{equation}
In particular, $x_1!=x_1= 1$, $x_2! = 2/(a+1)$. Also, $x_n = \varphi_{n-1}(a)/\varphi_{n}(a)$, and  
we  know from \cite{bergeronetal2012}  that $x_n \approx \sqrt{n/a}$ as $n \to \infty$.  
The corresponding polynomials and probability distributions are respectively  given by
\begin{equation}
\label{xnex2}
q_n(\eta) =  \frac{x_n!}{n!} \left(i\sqrt{\frac{a \eta}{2}}\right)^n\, H_n\left(-i\sqrt{\frac{\eta}{2a}} \right)\,  
\end{equation}
and
\begin{equation}
\label{hermprobdist}
\mathfrak{p}_k^{(n)}(\eta)= \eta^k (1-\eta)^{n-k} \frac{\varphi_k(a/\eta)\varphi_{n-k}(a/(1-\eta))}{\varphi_n(a)}\,. 
\end{equation}

\subsection{Asymptotic behavior at large $n$}
Let us evaluate the asymptotic behavior of the probability distribution \eqref{hermprobdist}. For that, let us rewrite it in terms of Hermite polynomials:
\begin{equation}
\label{hermprobdist2}
\mathfrak{p}_k^{(n)}(\eta)= \binom{n}{k}\eta^{\frac{k}{2}} (1-\eta)^{\frac{n-k}{2}} \frac{H_k\left(-i\sqrt{\frac{\eta}{2a}}\right)H_{n-k}\left(-i\sqrt{\frac{1-\eta}{2a}}\right)}{H_n\left(-i\sqrt{\frac{1}{2a}}\right)}\,. 
\end{equation}
Putting $k= nx$, with $0<x<1$, using the Stirling formula \eqref{stirlingA}, 
and the asymptotic behavior of Hermite polynomials  versus their respective degree when argument is not real \cite{magnus66}\footnote{Page 255. Actually, a factor 2 in front of $\left\vert H_n(t)\right\vert$  is missing there. }, 
\begin{equation}
\label{asHermite}
\left\vert H_n(t)\right\vert \sim \frac{n!}{2\,\Gamma\left(\frac{n}{2} + 1\right)}\, e^{\sqrt{2n}\vert \mathrm{Im}(t)\vert}\, , 
\end{equation}
we find
\begin{align}
\label{asympherm}
\nonumber \mathfrak{p}_{k=nx}^{(n)}&\sim \frac{1}{2}\,\binom{\frac{n}{2}}{\frac{nx}{2}} \eta^{\frac{k}{2}} (1-\eta)^{\frac{n-k}{2}} \exp\left\lbrack\sqrt{\frac{n}{a}}(\sqrt{x\eta}+ \sqrt{(1-x)(1-\eta)}-1)\right\rbrack\\
&\sim \frac{1}{2}\,\frac{1}{\sqrt{n\pi x(1-x)}}e^{n A(x)}\, , 
\end{align}
where 
\begin{equation}
\label{expoasherm}
A(x) = \frac{x}{2}\log\frac{\eta}{x} + \frac{(1-x)}{2}\log\frac{1-\eta}{1-x}+ \frac{1}{\sqrt{n a}}(\sqrt{x\eta} + \sqrt{(1-x)(1-\eta)} -1) \, . 
\end{equation}
Let us check if the asymptotic  distribution \eqref{asympherm}, continuous with respect to the measure $n\ud x$, is correctly normalized,
\begin{equation}
\label{nomrasherm}
\frac{1}{2}\,\sqrt{\frac{n}{\pi}}\int_0^1 [x(1-x)]^{-1/2}\, e^{n A(x)}\, \ud x = 1 \,?
\end{equation}
For showing this, we use Laplace's method. The two first derivatives of the function $A(x)$ are given by
\begin{align*}
 A^{\prime}(x)   & = \frac{1}{2}\left(\log\frac{\eta}{x} -   \log\frac{1-\eta}{1-x} \right) + \frac{1}{2\sqrt{n a}}\left(\sqrt\frac{\eta}{x} -   \sqrt\frac{1-\eta}{1-x} \right)\,  ,  \\
  A^{\prime\prime} (x) &  = - \frac{1}{2}\left(\frac{1}{x} + \frac{1}{1-x} \right) -\frac{1}{4\sqrt{n a}}\left(\sqrt\frac{\eta}{x^3} +   \sqrt\frac{1-\eta}{(1-x)^3} \right)\, . 
\end{align*}
We see that  in the integration interval $A^{\prime\prime} (x) < 0$, $A^{\prime}(x) = 0$ for $x=\eta$ (unique root), and that the values assumed by $A(x)$ and $A^{\prime\prime}(x)$ at this value are respectively
\begin{equation}
\label{AAeta}
A(\eta) = 0\, , \quad A^{\prime\prime}(\eta)\sim -\frac{1}{2\eta(1-\eta)}\, . 
\end{equation}
Then let us apply the Laplace approximation formula (with suitable conditions on the functions involved)
\begin{equation}
\label{laplaceappr}
\int_a^b h(x) \, e^{nA(x)}\, \ud x \sim \sqrt{\frac{2\pi}{n\vert A^{\prime\prime}(x_0)\vert}}\,h(x_0)\,e^{nA(x_0)}\,\ \mbox{as}\ n\to \infty\, , 
\end{equation}
where $A^{\prime}(x_0)= 0$ for $x_0\in [a,b]$, $A^{''}(x_0)< 0$  and $h$ is positive.
We get in our case,
\begin{equation}
\label{normhermasOK}
\frac{1}{2}\,\int_0^1\mathfrak{p}_k^{(n)}\, n\ud x \sim \sqrt{\frac{n}{\pi}}\int_0^1 [x(1-x)]^{-1/2}\, e^{n A(x)}\, \ud x \sim 1 \,. 
\end{equation}
\qed

\subsection{Boltzmann-Gibbs entropy}

>From the asymptotic behavior \eqref{asympherm} and \eqref{asymptbin} we infer the following behavior
\begin{equation}
\label{Hermasympprod}
\mathfrak{p}^{(n)}_{nx}\,\log\frac{\mathfrak{p}^{(n)}_{nx}}{\binom{n}{{nx}}}\sim \frac{1}{2\sqrt{n\pi}}\, h(x)\, e^{nA(x)}\, ,
\end{equation}
where $A(x)$ is given by \eqref{expoasherm} and the function $h(x)$ is given by
\begin{align}
\label{hxherm}
\nonumber h(x)= \sqrt{x(1-x)}&\left[ -\frac{1}{2}\log 2 + \frac{n}{2}[x\log(x\eta ) + (1-x) \log((1-x)(1-\eta)) ]+ \right.\\
 &\left. + \sqrt{\frac{n}{a}}[\sqrt{x\eta}+ \sqrt{(1-x-(1-\eta)}-1]\right] \, . 
\end{align}
After the usual replacement $\sum_{k=0}^n \mapsto \int_0^1n\ud x$, we get for the BG entropy, 
\begin{align}
\label{BGhermasymp}
\nonumber S_{\mathrm{BG}}&= - \sum_{k=0}^n \mathfrak{p}^{(n)}_k\,\log\frac{\mathfrak{p}^{(n)}_k}{\binom{n}{k}}\\
& \sim - \frac{1}{2}\sqrt{\frac{n}{\pi}} \int_0^1 h(x)\, e^{nA(x)}\, \ud x \, .
\end{align}
Applying the Laplace approximation method
\begin{align}
\label{BGhermasymplap}
\nonumber S_{\mathrm{BG}}&= - \frac{1}{2}\sqrt{\frac{n}{\pi}}\sqrt{\frac{2\pi}{n\vert A^{''}(\eta)\vert}}\,h(\eta)\,e^{nA(\eta)}\\
&\sim \frac{1}{2}\log 2 - n [ \eta\log\eta + (1-\eta)\log(1-\eta)]\,. 
\end{align}
So we can conclude that $S_{\mathrm{BG}}$ is extensive in this model. 

\subsection{Tsallis and R\'{e}nyi entropy}
To estimate the asymptotic behavior of both entropies, we first use the approximation resulting from \eqref{asympherm} and \eqref{asymptbin}
\begin{equation}
\label{asympttsare1}
\left[\binom{n}{k=nx}\right]^{q-1}\left(\mathfrak{p}_{k=nx}^{(n)}\right)^q \sim \frac{1}{\sqrt{2^{q+1}n\pi x(1-x)}}e^{n B(x)}\, , 
\end{equation}
with 
\begin{equation}
\label{defBx}
\begin{split}
B(x)&= q A(x) - (q-1) \mathcal{C}(x)
=\frac{q}{2}[x\log \eta + (1-x)\log (1-\eta)] + \\ &+ \left(\frac{q}{2} -1\right) [x\log x + (1-x)\log(1-x)]
+\\ &+ \frac{q}{\sqrt{na}}[\sqrt{x\eta}+ \sqrt{(1-x)(1-\eta)}-1]\, .
\end{split} 
\end{equation}
Next we transform the sum into an integral, as usual,
\begin{equation}
\label{asympttsare1}
\sum_{k=nx}\left[\binom{n}{k=nx}\right]^{q-1}\left(\mathfrak{p}_{k=nx}^{(n)}\right)^q \sim \sqrt{\frac{n}{2^{q+1}\pi}}\int_0^1 \ud x\, (x(1-x))^{-1/2}\, e^{n B(x)}\, . 
\end{equation}
In order to implement the Laplace method, we calculate $B^{\prime}$ and $B^{\prime \prime}$.
\begin{align}
\label{defBxp}
\nonumber B^{\prime}(x) &=\frac{q}{2}[\log \eta -\log (1-\eta)]  + \left(\frac{q}{2} -1\right) [\log x - (1-x)\log(1-x)]
+\\ &+ \frac{q}{2\sqrt{na}}\left[\sqrt{\frac{\eta}{x}}+ \sqrt{\frac{1-\eta}{1-x}}\right]\, ,\\
 \label{defBxpp} B^{\prime \prime}(x)&= \left(\frac{q}{2} -1\right)\,\frac{1}{x(1-x)}-\frac{q}{4\sqrt{na}}\left[\sqrt{\frac{\eta}{x^3}}+ \sqrt{\frac{1-\eta}{(1-x)^3}}\right]\,. 
\end{align}
We see that for $q< 2$ we have  $B^{\prime \prime}(x) < 0$ for all $x\in (0,1)$. 
Hence, if $q<2$ and if we find one and only one $x_0\in (0,1)$ such that $B^{\prime}(x_0)= 0$, the Laplace approximation method is valid, and we obtain the behavior of the sum at large $n$:
\begin{equation}
\label{hermsumas}
\sum_{k}\left[\binom{n}{k}\right]^{q-1}\left(\mathfrak{p}_{k}^{(n)}\right)^q \sim \sqrt{\frac{1}{2^{q}\vert B^{\prime \prime}(x_0)\vert}}\, (x_0(1-x_0))^{-1/2}\, e^{n B(x_0)}\, .
\end{equation}
Now, for the median value $\eta = 1/2$, we find  immediately the unique solution $x_0= 1/2$. Then, $B^{\prime \prime}(1/2)= 2(q-2) - q/\sqrt{na}$, $B(1/2) =(1-q)\log 2$, and so
\begin{equation}
\label{appsumhalf}
\sum_{k}\left[\binom{n}{k}\right]^{q-1}\left(\mathfrak{p}_{k}^{(n)}\right)^q \underset{\mbox{at large $n$}}{\sim} 2^{(3-q)/2}(q-2)^{-1/2}\, e^{n(1-q)\log 2}\, .
\end{equation}
Therefore, for $\eta= 1/2$, while the Tsallis entropy is not extensive, the R\'enyi entropy is extensive, 
\begin{equation}
\label{RenyiHermite1}
S_{\mathrm{Re};q} \sim  n \, \log 2 \, . 
\end{equation}
One can  easily show that with $\eta = 1/2 + \delta$, $\vert \delta \vert \ll 1/2$, the value of the root $x_0$ is $x_0 = \frac{q}{2-q}\delta + O(\delta^2)$ and that the behavior \eqref{RenyiHermite1} holds too. We have checked numerically that it holds for all $\eta \in (0,1)$. 
We notice that this behavior (which is simply $\sim n$ if we adopt the original R\'enyi choice $\log_2$) is the same as for the two other cases considered in this paper, Eqs.\,\eqref{Renyiqexp} and \eqref{RenyiAbel1}, and also for the binomial and Laplace de Finetti distributions considered in \cite{bergeronetal2014A}. We will come back to this important point in the conclusion. 

\section{Conclusions}

In this paper our main interest is the extensivity property of different entropies constructed from generalized binomial distributions. We analyse the behavior of three entropies, mainly the Boltzmann-Gibbs, Tsallis, and R\'enyi ones for the three examples of generalized binomial distributions presented in \cite{bergeronetal2013B}, recalling that our point of view is strictly microscopic. For that sake we examined the asymptotic behavior of the deformed probability distributions in question, which are those whose generating functions are the $q$-exponential, the exponential of the Lambert function and the exponential of a second-degree polynomial: the probabilities obtained are respectively the Polya distribution, a product of modified Abel polynomials and a product of Hermite polynomials. 
 
As could be expected, the Tsallis entropy is not extensive for the three probability distributions considered. The results found for the other two entropies are interesting: the R\'enyi entropy is extensive for the three probability distributions and, which is surprising, the Boltzmann-Gibbs one is extensive for two cases, those related to the $q$-exponential and to the Hermite polynomials, but not when the probability distribution is given by modified Abel polynomials.  This example of non-extensivity of Boltzmann-Gibbs is a result that deserves further investigation, as it has so far been considered as the universally extensive entropy.  As to the  R\'enyi entropy an important aspect of the result found here is that for all the three studied distributions its asymptotic value at large  $n$ is the same, $n \log 2$, and therefore does not depend on its parameter $q$.

Actually, this extensivity is probably due to the nature of the three distributions examined here, which are smooth deformations of the binomial one. We have shown in \cite{bergeronetal2014A} that both Boltzmann-Gibbs and R\'enyi are extensive for the binomial case.  Deformations of the binomial distribution introduce correlations, and these correlations may or not be strong enough to substantially modify the asymptotic behaviors. The fact that extensivity holds for R\'enyi and for its BG limit at $q=1$ when the deformed probability is either the Polya distribution or a product of Hermite polynomials indicates that in these cases the related correlations are weak. Otherwise,  the behavior of the deformed probability given as products of modified Abel polynomials is different as the Boltzmann-Gibbs limit of the R\'enyi entropy is asymptotically not extensive. This distribution deserves a further investigation on the correlations it introduces and we might expect them to be stronger than the two former mentioned cases; this issue will be the subject of  future work. Due to this exceptionality of the modified Abel polynomials case we illustrated it here with a concrete and non trivial probabilistic model.

\label{sec:conclus}

\appendix

\section{Axiomatic(s) for entropies}
\label{entropies}
As a complement to the introduction and since  the content of the paper is strongly concerned with  entropy,   we remind in this appendix, through different sets of postulates, the senses which can be given to this  mathematical entity. Entropy is at the same time an information theory concept and a physical quantity as well - physical in the sense that it should be accessible to measurement, and \emph{which acts}, according to Boltzmann,  \emph{as a link between the microscopic  and the macroscopic worlds}. 

 It is worthy to start with the way Shannon introduced it in \cite{shannon48}: 
\begin{quote}
We have represented a discrete information source as a Markoff process. Can we define a quantity which will measure, in some sense, how much information is ``produced” by such a process, or better, at what rate information is produced?
Suppose we have a set of $n$ possible events whose probabilities of occurrence are $p_1,p_2, \dotsc, p_n$. These probabilities are known but that is all we know concerning which event will occur. Can we find a measure of how much ``choice” is involved in the selection of the event or of how uncertain we are of the outcome?
If there is such a measure, say $H(p_1,p_2, \dotsc, p_n) $, it is reasonable to require of it the following properties:
\begin{enumerate}
  \item[S1] $H$ should be continuous in the $p_k$.
  \item[S2] If all the $p_k$ are equal, $p_k= 1/n$, then $H$ should be a monotonic increasing function of $n$. With equally $n$ likely events there is more choice, or uncertainty, when there are more possible events.
  \item[S3] If a choice be broken down into two successive choices, the original $H$ should be the weighted sum
of the individual values of $H$.
\end{enumerate}
Then (Theorem) the only $H$ satisfying the three above assumptions is of the form:
\begin{equation}
\label{shannon1}
H\equiv  H(\mathcal{P})=-K\sum_{k=1}^n p_k\log p_k \equiv \langle -K \log p_k\rangle\, . 
\end{equation}
where $K$ is a positive constant and $\mathcal{P}= (p_1,p_2, \dotsc,p_n)$. 
\end{quote}
The above identity means  that if we look at $k\mapsto  -K\log p_k = y_k$ as a random variable $Y$, then the entropy $H$ is its expected value with respect to the distribution $k\mapsto p_k$, $H= \langle Y\rangle$. Information theory uses $\log_2$ instead of $\log$ and \eqref{shannon1} with $K=1$ is the average number of bits needed to describe any random variable  with the same probability distribution. 

A (partially) different set of axioms, which involve conditional probabilities, was established by Khinchin \cite{khinchin53} in view of characterizing the Shannon entropy \eqref{shannon1}. Here we also use  the notation $H[\xi]\equiv H(\mathcal{P})$ where $\xi=(x_1,x_2,\dotsc,x_n)$ is a random variable with probability distribution $\mathcal{P}$, i.e. $p_k$ is the probability that $\xi$ assumes the value $x_k$.  
\begin{enumerate}
  \item[K1] $H$ is symmetrical in its arguments.
  \item[K2]  The uniform distribution $p_k = 1/n$ has maximal $H= K\log n$.
  \item[K3] If  $\mathcal{Q}= (q_1,q_2,\dotsc, q_m)$ is a probability distribution with $m>n$, $q_k=p_k$ for $1\leq k\leq n$ and $q_k=0$ for $n+1\leq k\leq m$, then $H(\mathcal{P})= H(\mathcal{Q})$. 
 \item[K4] For any random variables $\xi$ and $\eta$, $H[\xi,\eta] = H[\xi] + \sum_{k=1}^n p_k\,H[\eta \,|\, \xi= x_k]$, 
which means that the  joint entropy is the sum of the entropy of one variable, plus the average value of the entropy of the other variable, once the first is given.
\end{enumerate}

According to R\'enyi in \cite{renyi60}  the Shannon entropy is characterized by another  (partially different) set of  postulates (Fadeev):
\begin{enumerate}
  \item[F1=K1] $H$ is symmetrical in its arguments.
  \item[F2] $H(p,1-p)$ is a continuous function of $p$ for $0\leq p\leq1$. 
  \item[F3] $H(1/2,1/2)=1$. 
 \item[F4] $H(t p_1,(1-t)p_1, p_2, \dotsc, p_n) = H(p_1,p_2, \dotsc, p_n) + p_1H(t,1-t)$ for any distribution 
$\mathcal{P}= (p_1,p_2,\dotsc, p_n)$ and for $0\leq t\leq1$.
\end{enumerate}
The Shannon entropy is also the only one which satisfies these four postulates. On the other hand, there are many quantities other than \eqref{shannon1} that satisfy F1, F2 and F3, plus the property of additivity 
\begin{equation}
\label{addentr}
H(\mathcal{P}\ast\mathcal{Q}) = H(\mathcal{P}) + H(\mathcal{Q})\, , 
\end{equation}
where $\mathcal{P}\ast\mathcal{Q}$ is the direct product of the distributions $\mathcal{P}$ and $\mathcal{Q}$. The fundamental property \eqref{addentr} is weaker than the Shannon S3. R\'enyi in \cite{renyi60} gave the following example which now bears the name of \textit{R\'enyi entropy}:
\begin{equation}
\label{renyient0}
H_q(p_1,p_2,\dotsc, p_n)= \frac{1}{1-q}\log_2\left(\sum_{k=1}^n p_k^q\right)\equiv \log_2\left(\left\langle p_k^{q-1}\right\rangle\right)^{1/(1-q)}\, ,
\end{equation} 
where $q>0$ and $q\neq 1$, which is one of the  entropies  examined in this paper. Usually $\log_2$ is replaced by $\log$. This family of entropies goes to the Shannon entropy as $q\to 1$. 

To dispel any remnant ambiguity regarding the definition of both the above entropies if one wants to impose additivity,  R\'enyi  defined a set of 5 postulates that characterize completely  these quantities. First, he extended his considerations to \textit{incomplete} distributions, i.e. sequences $\mathcal{P}= (p_1,p_2, \dotsc,p_n)$ of non-negative  numbers such that their \textit{weights}
\begin{equation}
\label{weightdist}
\texttt{w}(\mathcal{P}) := \sum_{k=1}^n p_k 
\end{equation}
are positive and $\leq 1$, but not necessarily equal to 1.  The R\'enyi  postulates for the entropy function $H(\mathcal{P})$ are
\begin{enumerate}
  \item[R1] $H(\mathcal{P})$ is a symmetric  functions of the elements of $\mathcal{P}$.
  \item[R2] If $\{p\}$ denotes the generalized probability distribution consisting of the single probability $p$ then $H(\{p\})$ is a continuous function of $p$ for $0 < p \leq1$ (not necessarily in $0$).
  \item[R3] $H(\{1/2\})=1$. 
 \item[R4] Additivity holds for any pair of incomplete distributions, $H(\mathcal{P}\ast\mathcal{Q}) = H(\mathcal{P}) + H(\mathcal{Q})
 $.
  \item[R5] There exists a strictly monotonic and continuous function $y=g(x)$ such that for two incomplete distributions $\mathcal{P} = (p_1,p_2, \dotsc,p_m)$ and \linebreak  $\mathcal{Q} = (q_1,q_2, \dotsc,q_n)$   with $\texttt{w}(\mathcal{P}) + \texttt{w}(\mathcal{Q}) \leq 1$, we have the  $g$-mean value formula
\begin{equation}
\label{extmeanH}
H(\mathcal{P} \cup \mathcal{Q}) = g^{-1}\left[ \frac{\texttt{w}(\mathcal{P})\,g\left(H(\mathcal{P})\right)+ \texttt{w}(\mathcal{Q})\,g\left(H(\mathcal{Q})\right)}{\texttt{w}(\mathcal{P}) + \texttt{w}(\mathcal{Q})}\right]\, . 
\end{equation}
\end{enumerate}
By adding some considerations involving conditional probability, R\'enyi proved that there are only two possible solutions for the function $g$.
\begin{itemize}
  \item The function $g$ is linear, $g(x) = a x +b$, and then the corresponding entropy is Shannon for incomplete distributions
  \begin{equation}
\label{shannoninc}
H(\mathcal{P})= - \frac{\sum_{k=1}^n p_k\log_2 p_k}{\sum_{k=1}^n p_k}\, . 
\end{equation}
  \item It is exponential, $g_q(x) = 2^{(q-1)x}$, $q> 0$, $q\neq 1$, and then the entropy is R\'enyi for incomplete distributions
  \begin{equation}
\label{renyiinc}
H_q(\mathcal{P})= \frac{1}{1-q}\log_2\left[ \frac{\sum_{k=1}^n p_k^q}{\sum_{k=1}^n p_k}\right]\, .
\end{equation}
\end{itemize}
The first case is the limit as $q \to 1$ of the second one.

Finally, we have as well considered the Tsallis entropy which is also a deformation of \eqref{shannon1}:
\begin{equation}
\label{renyient0}
S_q(p_1,p_2,\dotsc, p_n)= \frac{1}{q-1}\left(1- \sum_{k=1}^n p_k^q\right)\equiv  \left\langle \frac{1-p_k^{q-1}}{q-1}\right\rangle\, .
\end{equation} 
This entropy also goes to the Shannon entropy as $q\to 1$.  While it satisfies F1 and F2, the Tsallis entropy does not satisfies F3 and has the deformed additivity property
\begin{equation}
\label{defaddTs}
S_q(\mathcal{P}\ast\mathcal{Q}) = S_q(\mathcal{P}) + S_q(\mathcal{Q}) + (1-q) S_q(\mathcal{P})\,S_q(\mathcal{Q}) \, . 
\end{equation}
More precisely, Abe \cite{abe00} has proved that this entropy is characterized by three postulates  adapted from the Shannon-Khinchin axioms.
\begin{enumerate}
  \item[A1] $S_q(\mathcal{P})$ is continuous with 
respect to all its arguments and takes its maximum for the equiprobability distribution $p_k= 1/n$.
  \item[A2] If  $\mathcal{Q}= (q_1,q_2,\dotsc, q_m)$ is a probability distribution with $m>n$, $q_k=p_k$ for $1\leq k\leq n$ and $q_k=0$ for $n+1\leq k\leq m$, then $S_q(\mathcal{P})= S_q(\mathcal{Q})$. 
   \item[A3]  For any random variables $\xi$ and $\eta$, $$S_q[\xi,\eta] = S_q[\xi] + S_q[\eta\,|\, \xi]  + (1-q) S_q[\xi]\,  S_q[\eta\,|\, \xi] \,.$$ 
\end{enumerate}

\section{Asymptotic formulas}
\label{asymptfor}
From the Stirling formula, 
\begin{equation}
\label{stirlingA}
n!\sim\sqrt{2\pi} \,e^{-n}\,n^{n+1/2}\quad \mbox{at large $n$}\,,
\end{equation}
we derive the asymptotic behavior of binomial coefficient at large $n$,
\begin{equation}
\label{asymptbin}
 \binom{n}{k=nx} \sim \frac{1}{\sqrt{2\pi nx(1-x)}}\, e^{n\mathcal{C}(x)}
\end{equation}
where the function $\mathcal{C}(x):=- x\log x - (1-x)\log(1-x)$.
From this expression, we can check that the summation formula 
\begin{equation}
\label{binsum}
\sum_{k=0}^n\binom{n}{k} = 2^n 
\end{equation}
keeps its validity at large $n$. Indeed, with $k=nx$, $0<x<1$ and  replacing  the  above sum $\sum_{k=0}^n$ by the integral $\int_0^1 n\ud x$, leads to
\begin{equation}
\label{binsumint}
\sum_{k=0}^n\binom{n}{k} \sim \sqrt{\frac{n}{2\pi}}\int_{0}^1[x(1-x)]^{-1/2}\, e^{n\mathcal{C}(x)}\, \ud x\,. 
\end{equation}
Then we apply the Laplace's method for evaluating the above integral.  Laplace's approximation formula (with suitable conditions on the functions involved) reads 
\begin{equation}
\label{laplaceappr}
\int_a^b h(x) \, e^{nA(x)}\, \ud x \approx \sqrt{\frac{2\pi}{n\vert A^{''}(x_0)\vert}}\,h(x_0)\,e^{nA(x_0)}\,\ \mbox{as}\ n\to \infty\, , 
\end{equation}
where $A^{'}(x_0)= 0$ for $x_0\in [a,b]$, $A^{''}(x_0)< 0$  and $h$ is positive.
Here, we have 
\begin{equation*}
\mathcal{C}^{\prime}(x)= -\log(x) +\log(1- x)\, \quad \mathcal{C}{^{\prime\prime}}(x) = - \frac{1}{x(1-x)}\, .
\end{equation*}
Now, $x=1/2$ is the only root of $\mathcal{C}^{\prime}(x) = 0$ in the interval $0<x<1$, and this corresponds to the maximum of  $\mathcal{C}(x)$ in that interval: $\mathcal{C}(1/2) = \log 2$. Moreover, $\mathcal{C}{^{\prime\prime}}(1/2) = 4$. Thus, 
\begin{equation*}
\label{binsumint}
 \sqrt{\frac{n}{2\pi}}\int_{0}^1[x(1-x)]^{-1/2}\, e^{n\mathcal{C}(x)}\, \ud x \sim 2^n\,. 
\end{equation*}

\section*{Acknowledgments}
H. Bergeron and J.P. Gazeau thanks the CBPF and the CNPq for financial support and CBPF for hospitality. E.M.F. Curado acknowledges CNPq and FAPERJ for financial support.

\end{document}